\documentclass[preprintnumbers]{revtex4}
\usepackage{graphicx}
\usepackage{amsmath}
\usepackage{bm}
\newcommand{\pd}[2]{\frac{\partial {#1}}{\partial {#2}}} % \pd ft gives partial f partial x.

\newcommand{\ga}{{\gamma}}

\newcommand{\de}{{\delta}}
\newcommand{\De}{{\Delta}}

\newcommand{\ka}{{\kappa}}

\newcommand{\om}{{\omega}}
\newcommand{\Om}{{\Omega}}

\begin{document}
\title{Storage of light: A useful concept?}
\author{M.G. Payne}
\affiliation{Georgia Southern University, Statesboro, GA 30460-8031}
\author{L. Deng}
\affiliation{Electron and Optical Physics Division, NIST, Gaithersburg, MD 20899-8410}
\date{\today}
\begin{abstract}
We show both analytically
and numerically that photons from a probe pulse are not stored in several
recent experiments.  Rather, they are absorbed to produce a two-photon
excitation.  More importantly, when an identical coupling pulse is re-injected into
the medium, we show that the regenerated optical field has a pulse width that is
very different from the original probe field. It is therefore, not a faithful copy of the original probe pulse.
\end{abstract}
\maketitle

\vskip 20pt
Recent experiments [1,2] on the extremely slow propagation of a pair of optical pulses in
a resonant medium and the subsequent coherent regeneration of the ``probe"
pulse have generated much excitement in the fields of optical computing and quantum optics.
The physical explanation offered to these extraordinary
phenomena rests on the adiabatic theory developed a decade ago by Eberly and
coworkers[3]. To a large degree, the ``dark-state polariton" theory [4] and
all other recent works formulated around it are basically a field theoretical reformulation
of the adiabatic approximation of pulse pair propagation in
resonant media originally studied reasonably complete in the early 90's.
These new treatments have
provided some further understanding of the process, especially from the view
point of atomic spin wave excitation and atomic coherence.
Two main results of the these works, however, do not, in our view, accurately
describe the physical picture of such a highly nonlinear excitation process.
In this letter, we show, based on the adiabatic theory, that the claim of
the ``revival of the probe pulse" in recent experiments [1,2,5] is not
accurate.  Indeed, we show that under the conditions of these experiments
the original probe photons are neither ``stopped" or ``stored".  Rather,
they are
absorbed to produce two-photon excitation.
Furthermore, we show that the `` retrieval of the probe pulse"
is not achieved without restrictions being placed on the characteristics of
the second coupling pulse used to regenerate the ``probe" field.  In fact, we show that
the regenerated field has time-spectra characteristics that are very different
from the original probe field.
Therefore, ``storage of light" is not an accurate description
of the process.  In addition, we show that all experimental observables
can be well predicted with the
usual treatment of combining classical electrodynamics and the three-state
model without invoking a full field theoretical methodology. This is not
surprising since when a detailed handling of spontaneous emission is not
critical and the fields are not extremely weak, a mean-field approximation
to the quantum treatment of the electromagnetic field is valid.  Our analysis
and results on the physical picture are unique and provide a clear understandings
of the physical process.  To the best of our knowledge the subtle understandings
presented here on the ``tapped light" problem are not contained in
either the original studies or any of the subsequent studies, including all recent
theoretical works, on the subject.

\vskip 10pt
We start with a $\Lambda$\ system (Fig. 1) with the assumption that
a probe laser ($E_p$, frequency $\om_p$)
is tuned on or near resonance with the $|1>\rightarrow |2>$\ transition.  In addition,
a coupling laser ($E_c$, frequency $\om_c$) is tuned so that exact two-photon
resonance between states $|1>$ and $|3>$ is achieved.  We therefore have the following
atomic equations of motion
\begin{subequations}
\begin{eqnarray}
\pd{A_1}{t_r}&=&i\Om_{p}A_2,\\ 
\pd{A_2}{t_r}&=&i\Om_{p}^{*}A_1+i\Om_{c}^{*}A_3+i\left(\delta+i\frac{\ga_2}{2}\right)A_2,\\ 
\pd{A_3}{t_r}&=&i\Om_{c}A_2,
\end{eqnarray}
\end{subequations}
where $A_j$ and $\ga_j$ are the $jth$ amplitude of the atomic wave function and decay
rate, respectively, and we have made phase transformation to remove $z$ dependent
phase factors.  Also, $\Om_{p(c)}^{*}=D_{21(23)}E_{p(c)}/(2\hbar)$,
$\de=\om_p-\om_{21}$, $\om_p-\om_c=\om_{31}$, and
$t_r=t-z/c$\ is the retarded time which, in vacuum, is the
combination of $t$ and $z$ that the laser field amplitudes depend on.  Notice that
we have included non-vanishing one-photon detuning $0\le\de<\Om_c$, a feature that is not included
in both the original adiabatic theory and those developed recently around
the ``dark-state polariton" theory.

\vskip 10pt
In order to correctly predict the propagation
of the probe and coupling laser pulses in a resonant medium, Eq.(1) must be
solved simultaneously with Maxwell equations describing
the propagation of the probe and coupling fields.  For plane
waves in the slowly varying amplitude approximation the Maxwell equations
resume the form
\begin{subequations}
\begin{eqnarray}
\left(\pd{\Om_{p}^{*}}{z}\right)_{t_r}&=&i\kappa_{12}A_{1}^{*}A_2,\\ 
\left(\pd{\Om_{c}^{*}}{z}\right)_{t_r}&=&i\kappa_{32}A_{3}^{*}A_2,
\end{eqnarray}
\end{subequations}
where $\kappa_{12(32)}=2{\pi}N\om_{p(c)}|D_{12(32)}|^2/(\hbar c)$.
The process involves first seeking, from Eq.(1), an adiabatic solution
for the atomic response, and
applying the result to Eq.(2) to self-consistently describe the
propagation of the pulse pair within the adiabatic limit.
Following this procedure, we thus obtain adiabatic solution to the atomic response [3]
\begin{subequations}
\begin{eqnarray}
A_1(z,t_r)&=&\frac{\Om_c^{*}(z,t_r)}{\Om(z,t_r)},\\ 
A_2(z,t_r)&=&-\frac{i}{\Om_p}\pd{A_1}{t_r}=-\frac{i}{\Om_c}\pd{A_3}{t_r},\\ 
A_3(z,t_r)&=&-\frac{\Om_p^{*}(z,t_r)}{\Om(z,t_r)}.
\end{eqnarray}
\end{subequations}
where $\Om(z,t_r)=\sqrt{|\Om_{p}(z,t_r)|^2+|\Om_{c}(z,t_r)|^2}$.  As usual,
with the adiabatic approximation used in a resonance situation, we must
have $\Om_c$\ already strong when $\Om_p$\ starts to build up.  One scenario where
the adiabatic approximation would hold through the whole laser pulse is
if the two lasers peak at the same time, but the pulse length of the coupling laser
is much longer. This feature of the pulse lengths and the requirement $|\Om_c\tau|>>1$\ through
the pulse length, $\tau$, of the probe laser will make results based on the adiabatic
approximation quite accurate.

\vskip 10pt
With Eq.(3) as atomic response, we now solve Eq.(2) that now resumes the form of
\begin{equation}
\left(\pd{\Om_{p(c)}^{*}}{z}\right)_{t_r}=-\frac{\ka_{12(32)}\tau}{\sqrt{|\tau\Om_c|^2+|\tau\Om_p|^2}}\frac{\partial}{\partial t_r/\tau}\left(\frac{\tau\Om_{p(c)}^{*}}{\sqrt{|\tau\Om_c|^2+|\tau\Om_p|^2}}\right).
\end{equation}
We first note that the quantity
\begin{equation}
F(z,t_r)=\frac{|\Om_p|^2}{\ka_{12}}+\frac{|\Om_c|^2}{\ka_{32}}
\end{equation}
represents the sum of the photon fluxes at $\om_p$\ and $\om_c$\ divided by
the concentration of the medium through which the waves propagate. Differentiate
$F$\ with respect to $z$ while holding $t_r$\ fixed and apply Eq.(4), we immediately
reach the conclusion that $F$\ depends only on $t_r$.  This
permits one to evaluate $F$\ by evaluating it at $z=0$, the entrance to the atomic
vapor cell, e.g.
\begin{equation}
F(z=0,t)=\frac{|\Om_c(0,t)|^2}{\ka_{32}}+\frac{|\Om_p(0,t)|^2}{\ka_{12}}.
\end{equation}
Therefore, whenever $F(z,t_r)$\ occurs
it can be replaced by $F(0,t_r)$, as determined in Eq.(6)[6]. The lack of dependence of this
quantity on $z$ when the full set of equations is solved numerically is an important
test for the validity of the adiabatic approximation.

\vskip 10pt
Let us consider, for simplicity, the case where $\ka_{12}=\ka_{32}$.  Since
$\sqrt{|\Om_c|^2+|\Om_p|^2}$\ does not depend on $z$ when $t_r$\ is held
fixed, Eq.(4) can be recast into
\begin{equation}
\frac{\partial}{\partial z}\left(\frac{\Om_{p(c)}^{*}\tau}{\sqrt{|\Om_c\tau|^2+|\Om_p\tau|^2}}\right)=-\frac{\ka_{12(32)}\tau}{|\Om_c\tau|^2+|\Om_p\tau|^2}\frac{\partial}{\partial t_r/\tau}\left(\frac{\Om_{p(c)}^{*}\tau}{\sqrt{|\Om_c\tau|^2+|\Om_p\tau|^2}
}\right).
\end{equation}
Define
\begin{subequations}
\begin{eqnarray}
W_p&=&\frac{\Om_p^{*}\tau}{\sqrt{|\Om_c\tau|^2+|\Om_p\tau|^2}},  W_c=\frac{\Om_c^{*}\tau}{\sqrt{|\Om_c\tau|^2+|\Om_p\tau|^2}},\\
v(t_r)&=&\int_{-\infty}^{t_r/\tau}\left(|\Om_{c}(0,t_r^{'})\tau|^{2}+|\Om_{p}(0,t_r^{'})\tau|^{2}\right)d\left(\frac{t_r^{'}}{\tau}\right),  u(z)=\int_{0}^z\ka_{12}\tau dz^{'},
\end{eqnarray}
\end{subequations}
Equation (7) now becomes
\begin{subequations}
\begin{eqnarray}
\pd{W_p}{u}+\pd{W_p}{v}&=&0,\\ 
\pd{W_c}{u}+\pd{W_c}{v}&=&0,
\end{eqnarray}
\end{subequations}
where general ``travelling wave" type solutions are immediately obtained as
\begin{subequations}
\begin{eqnarray}
{W_p}&=&F_p(v-u),\\ 
{W_c}&=&F_c(v-u).
\end{eqnarray}
\end{subequations}
The functions $F_p$\ and $F_c$\ are easily determined by evaluating at $z=0$ (so $u=0$).
When the second coupling laser pulse is injected into the medium after a time
delay, the predictions about the revival of the ``probe" pulse are all contained
in this solution.  The tabulation of $F_p$\ remains the same as long as only
a second coupling pulse is sent into the medium at a later time.  Notice that
if $|\Om_{p}(z,t_r)|^2<<|\Om_{c}(z,t_r)|^2$, the population in state
$|3>$\ is always small and the coupling laser propagates as in vacuum.  In this
limit, with the replacement of $\Om_{c}(z,t_r)=\Om_{c}(0,t_r)$, Eq. (10) can
still be used to determine $\Om_{p}(z,t_r)$\ even for $\ka_{12}\ne\ka_{32}$.
This is the essence of the approximations applied in the
works by Ref.[4].

\vskip 10pt
We now examine the wave propagation and ``probe revival" with the field profiles [7]
\begin{subequations}
\begin{eqnarray}
\Om_{p}(0,t)&=&\Om_{p0}e^{-(\frac{t}{\tau})^2},\\
\Om_{c}(0,t)&=&\Om_{c0}\left(e^{-0.2(\frac{t}{\tau})^2}+Re^{-0.2(\frac{t}{\tau}-x_0)^2}\right).
\end{eqnarray}
\end{subequations}
Here, $\Om_{p0}$\ and $\Om_{c0}$\ are real constants characterizing the
peak amplitudes of the two half-Rabi frequencies before the pulses enter the resonant
medium,
$R$\ is the ratio of the Rabi frequency at which the coupling laser recurs to its initial
amplitude, and $x_0=t_d/\tau$\ is the value of $t_r/\tau$\ at which the peak of the coupling
laser recurs.

\vskip 10pt
We first consider the case where
$R=0$ and the group velocity of the probe
pulse is sufficiently small so that the coupling laser dies away before the
probe pulse can propagate through the
cell.  When the coupling laser begins to die out, the intensity of the two lasers becomes
proportional to each other, therefore the relation
\begin{equation}
v(t_r)\simeq |\Omega_{c0}\tau|^2\sqrt{5\pi/2}+|\Omega_{p0}\tau|^2\sqrt{\pi/2}
\end{equation}
is appropriate.  Also, $|\Omega_{p}(0,t_r)|^2$\ has long been very small compared with $|\Om_{c}(0,t_r)|^2$.
Thus, for $t_r/\tau>3.0$\ we have, as a very good approximation
\begin{equation}
\Om_{p}^{*}(z,t_r)\simeq \Om(0,t_r)F_p((|\Om_{c0}\tau|^2\sqrt{5\pi/2}+|\Om_{p0}\tau|^2\sqrt{\pi/2})-\ka_{12}\tau z).
\end{equation}
At such a late time during the pulse, the argument of $F_p$\ depends
only on the $z$ coordinate so that the time dependence of
$\Om_{p}^{*}$\ is obviously exactly the same as that of $\Om(0,t_r)$.
When $|\Om_c(0,t_r)|$\ is several times larger than $|\Om_p(0,t_r)|$,
as always is at such
late times at $z=0$, this means that $\Om_{p}^{*}(z,t_r)$\ has the same time dependence
as $\Om_{c}^{*}(0,t_r)$\ at such late times. Note that at late times
$\Om_{p}^{*}(z,t_r)/\Om(z,t_r)=A_3(z,t_r)$ is also
independent of retarded time, indicating that as $\Om_{p}(z,t_r)$\ approaches
zero, the ratio of populations in $|1>$\ and $|3>$\ stays fixed.  It is precisely
this way of having two fields to go to zero with fixed ratio that preserves the
the adiabatic approximation during the process.  There must be a coherent superposition of states
in $|1>$\ and $|3>$, such that $A_3(z,t_r)/A_1(z,t_r)=\Om_{p}^{*}(z,t_r)/\Om_{c}^{*}(z,t_r)$.
A similar relation holds when the second coupling laser pulse starts to
build up.  In order for the behavior to remain adiabatic, the probe laser
half-Rabi frequency must build up proportional to the coupling laser, with
the ratio of the two being the local ratio of $A_3/A_1$.  This persists until
the depletion of the population of $|3>$\ forces the ratio to decrease at later
$t_r$. By choosing $z$ such that $2\ka_{12}\tau z=|\Om_{c0}\tau|^2\sqrt{5\pi/2}+\sqrt{\pi/2}|\Om_{p0}\tau|^2$,
we thus have the same argument of $F_p$\ at $z=0$\ and $t_r=t=0$, resulting
$F_p=\Om_{p0}/\sqrt{|\Om_{c0}|^2+|\Om_{p0}|^2}$.
This is the largest value that $F_p$\ takes on, and at this depth into
the medium the value of $|A_3(z,t_r)|$\ matches its largest value at $z=0$.
This population persists at large $t_r$\ until very slow collisional effects
destroy the coherence left behind in states $|1>$\ and $|3>$.
It is this long persistence of
a coherent mix of populations in states $|1>$\ and $|3>$\ that leads to the
regeneration of an optical field when a delayed second coupling pulse is injected
into the medium. Of course, the number of atoms left in $|3>$\ is (within the
adiabatic approximation) equal to the number of photons in the original probe pulse.

\vskip 10pt
We now investigate the case where $\Om_{c0}\tau=20$, $\Om_{p0}\tau=5$,
$\ga_2\tau=0$, $\kappa_{12}\tau=\ka_{32}\tau=200 cm^{-1}$, and $R=4$.  Based on our
adiabatic theory, we predict a maximum population of $|A_3|=1/\sqrt{17}$\ at
a depth of 2.86 cm into the medium.  This is indicated
by the line of constant color in a contour plot
of $|A_3(z,t_r)|$\ (see Fig. 2a) leading from $t_r=0$\ and $z=0$\ out to the
horizontal path at $z\simeq 2.86$\ cm.  A corresponding surface plot for
$\Omega_{p}\tau$\ as functions of $t_r/\tau$\ at $z=3$\ cm is given in Fig. 3a
which shows the long asymmetric tail on $\Omega_p$\
as described above. In the region between $2.5\le{t_r}/\tau\le{5}$ the
ratio of the two half-Rabi frequencies is close to constant, averaging
around 0.24.  This ratio is also close to the
adiabatic approximation for $A_3$ since $|\Om_{p}|^2<<|\Om_{c}|^2$.  In the same
region, however, there is no optical field left, indicating no photons are ``stored"
or ``stopped".  Every probe photon is converted to the excitation of the state $|3>$ [8].

\vskip 10pt
We now come to an important point, e.g. to show that the regenerated
field has characteristics that are very different from that of the original
probe pulse.  That is, the regenerated
field is not a faithful copy of the original probe pulse unless
careful restrictions are placed on characteristics of the
second coupling laser pulse.  Recall
that $F_p$\ was determined from the functional dependence of the probe and coupling
laser pulses at $z=0$, in particular, $W_p(v(t))=-A_3(0,t)$, we thus have,
during the recurring coupling laser pulse
\begin{equation}
v(t_r)=S+\frac{R^2}{2}|\Om_{c0}\tau|^2\sqrt{5\pi/2}\left(1+erf\left(\sqrt{2/5}\frac{(t_r-t_d)}{\tau}\right)\right),
\end{equation}
where $S=|\Om_{c0}\tau|^2\sqrt{5\pi/2}+|\Om_{p0}\tau|^2\sqrt{\pi/2}$.
Equation (14) is of central importance in the following analysis on
the characteristics of the regenerated field.
We first note that $R$\ must be large enough so that $v(t_r)-\ka_{12}\tau z_m>0$.  That is,
$R$ must be large enough so that the group velocity of the regenerated photons
are large enough to exit the cell before the laser
induced transparency ends. Three time markers, therefore, are important
for
describing the regenerated field when it reaches the end of cell where $z=z_m$.
The first marker is the earliest time at which $F_p(0)=0$.
This time marker is determined by $v(t_{r1})-\ka_{12}\tau z_m=0$.
At a later time the value of $F_p(S/2)$\ will be equal to $-A_3(0,0)$.  This
second marker represents the time when $F_p$, for the given set of
parameters, reaches it maximum value at $z_m$, and it is determined by $v(t_{rm})-\ka_{12}\tau z_m=S/2$.
Finally, the third marker is the time at which the regenerated pulse
completes its exit from the cell, i.e. the time at which $F_p(S)=0$
and $v(t_{r2})-\ka_{12}\tau z_m=S$.  When these relations are used in
Eq.(14), we immediately obtain
\begin{equation}
\Om_{p}^{*}(z_m,t_{rm})=R\Om_{c0}e^{-(t_{rm}-t_d)^2/(5\tau^2)}\frac{\Om_{p0}\tau}{\sqrt{|\Om_{c0}\tau|^2+|\Om_{p0}\tau|^2}}.
\end{equation}
Notice that if $R$\ is chosen to make the argument of $F_p$\ at the time $t_r=t_d$
exactly the same as its value at $z=0$ and $t_r=t=0$, we then have
\begin{equation}
|\Om_{p0}\tau|^2\sqrt{\pi/8}+|\Om_{c0}\tau|^2\sqrt{5\pi/8}(1+R^2)=\ka_{12}\tau z_m,
\end{equation}
and
\begin{equation}
\Om_{p0}^{*}(z_m,t_d)=R|\Om_{c0}|\frac{\Om_{p0}\tau}{\sqrt{|\Om_{c0}\tau|^2+|\Om_{p0}\tau|^2}}.
\end{equation}
By determining the value of $t_r$\ such that $v(t_r)-\ka_{12}\tau z_m = 0$ or $S$,
a range of time over which the regenerated pulse rises from zero and returns to zero
at the exit of the cell may be obtained.
We thus estimate the FWHM pulse length, in the unit of the original probe pulse
length $\tau$, to be
\begin{equation}
\De_{1/2}=\frac{\sqrt{5\pi/2}}{R^2}\left(1+\frac{1}{\sqrt{5}}\left|\frac{\Om_{p0}}{\Om_{c0}}\right|^2\right).
\end{equation}
The key results shown in Eqs.(14-18) indicate that the regenerated field
is not a replica of the original probe pulse.  Therefore,
the concept of ``storage of light", in the reported experimental studies [1,2], is not accurate.
Furthermore, Eq.(15)
clearly indicates that the instantaneous phase of the regenerated field
is closely related to that of the second coupling pulse, in additional to the
phase of the atomic coherence $\rho_{32}$, therefore, cannot
be just that of the original probe pulse
alone.  In fact, if the probe laser has only penetrated a small fraction of the thickness of the
vapor cell when the coupling laser pulse has passed by, then $R$\ will turn out
to be much larger than unity. This means that  the width of the regenerated
field is generally much smaller than the width of the initial probe pulse.
Correspondingly, the bandwidth of the former will be much larger as can be seen
from Fig. 3a.  Both Fig. 2a and 3a show that in ``reviving the probe pulse"
every regenerated photon comes at the expense of flipping
population from state $|3>$\ to $|1>$. When the population of $|3>$\ has
been exhausted, there can be no further photon generated.

\vskip 10pt
Extensive numerical calculations carried out by
simultaneously solving Eqs.(1-2) have shown very good agreement with the
adiabatic calculation described above.  In Figs. 2b and 3b we show a contour
plot of $|A_3(z,t_r)|$ and a surface plot of $\Om_{p}^{*}$
for the same parameters given in Figs. 2a and 3a.  From Fig. 3b, we notice that
there is no probe field between $t_r/\tau=2$\ and the arrival time of the second
coupling pulse, i.e. there is no photon ``left" or ``stored" in the medium.
All probe photons have been absorbed in producing the coherent excitation of
the state $|3>$ as can be seen from Fig. 2b.  The case with
non-vanishing one-photon detuning in our adiabatic theory also produces a
result that is in very good agreement with numerical calculations.  Numerical
calculations have also been vigorously tested by making use of the fact that
with $\ga_2\tau=0$, the sum of the squares of the three state amplitudes
should be unity.  In all numerical examples described in this paper the
condition of unity was preserved through at least seven significant figures
if $\ga_2\tau=0$.

\vskip 10pt
We have shown analytically and numerically that
when $0\le\de<|\Om_c|$, a coherent optical
field with the frequency very close to that of the original probe
can be regenerated by re-injecting a coupling pulse.
Detailed analysis
on the conditions and characteristics of this regenerated field, including
the estimate of the pulse width, has shown that it is not the replica
of the original probe pulse.  We, therefore, caution the use of the concept
of ``storage of light" in the context of recently reported experiments, since
there has no evidence that this is indeed the case [9].  In fact, one can couple
the states $|1>$ and $|3>$ with a coherent magnetic pulse to create the
coherence required.  Under this circumstance, if an optical pulse that
couples the states $|2>$ and $|3>$, commonly referred to as the coupling pulse,
is injected into the system, an optical pulse that couples the states $|2>$
and $|1>$, commonly referred to as the probe pulse, will be generated.  The
latter is most certainly not the replica of the magnetic pulse used
to create the coherence.

\vskip 20pt
\centerline{References}
\vskip 10pt
\begin{itemize}
\item[1.] {C. Liu et al., Nature (London) 409, 490 (2001).}
\item[2.] {D.F. Phillips et al.,Phys. Rev. Lett. 86, 783 (2001).}
\item[3.] {J. Oreg, F.T. Hioe and J.H. Eberly, Phys. Rev A29,690 (1984); J.R. Kuklinski,
U. Gaubatz, F.T. Hioe, and K. Bergmann, Phys. Rev. A 40 R6749 (1989).  See also,
R. Grobe and J.H. Eberly, Laser Physics 5, 542 (1995); J.H. Eberly,
A. Rahman, and R. Grobe, Laser Physics 6, 69 (1996); J.R. Csesznegi and
R. Grobe, Phys. Rev. Lett. 79, 3162 (1997).}
\item[4.] {M. Fleischhauer, Opt. Express 4, 107 (1999); M. Fleischhauer, S.F. Yelin,
and M.D. Lukin, Opt. Communi. 179, 395 (1999); M. Fleischhauer and M.D. Lukin,
Phys. Rev. Lett. 84, 5094 (2000).}
\item[.5] {E.A. Cornell, Nature (London) 409, 461 (2001).}
\item[6.] {Equation (7) can also be derived exactly from Eq.(3) for the case of slowly
varying $|A_2|^2$ and not too large $\ga_2$.}
\item[7.] {These field profiles are reasonable description of the experiments reported.
In fact, as long as the adiabatic conditions are observed different profiles will
lead to the similar conclusions.  In addition, numerical simulations have suggested
that even in the region where the adiabatic conditions are in valid, a similar conclusion
can still be reached, provided certain non-adiabatic restriction are observed.}
\item[8.] {Figure 3a shows that all probe photons are converted to the excitation
of the state $|3>$.  If one can regenerate, from the excitation of the state
$|3>$, photons with optical properties identical to that of the original
probe filed, then it would be correct to claim the revival of the probe field.
As we will show later, however, this is not the case in recent reported experiments.}
\item[9.] {We emphasize that the purpose of the present study is to show that
photons from the probe pulse are neither ``stopped" or ``stored" in the recently
reported experiments [1,2].  Our intension is to stimulate further studies for
truly storage of the quantum information.}
\end{itemize}

\vskip 20pt
\centerline{Figure captions}
\vskip 10pt

{Figure 1.  Energy level diagram showing relevant laser excitations.}

\vskip 10pt
Figure 2.  Contour plot of $A_3(z,t)$. a. Adiabatic solution. b. Full numerical solution.
Parameters
used: $\Om_p\tau=5$, $\Om_c\tau=20$, $\ga_2\tau=0$, $\kappa_{12}\tau=\kappa_{32}\tau=200 cm^{-1}$,
$R=4$, $t_d/\tau=11$.

\vskip 10pt
Figure 3.  Surface plot of $\Om_p(z,t)$. a. Adiabatic solution. b. Full numerical solution.
Parameters
used: $\Om_p\tau=5$, $\Om_c\tau=20$, $\ga_2\tau=0$, $\kappa_{12}\tau=\kappa_{32}\tau=200 cm^{-1}$,
$R=4$, $t_d/\tau=11$.

\end{document}